\documentclass[11pt]{amsart}
 
\usepackage{amssymb}
\usepackage{amsmath}
\usepackage{amsfonts}
\usepackage{amsthm}
\usepackage{graphicx}
\usepackage{epsfig}
\usepackage{longtable}

\newtheorem{thm}{Theorem}[section]
\newtheorem{lem}[thm]{Lemma}

\newtheorem{prop}[thm]{Proposition}

\theoremstyle{definition}
\newtheorem{defn}[thm]{Definition}

\theoremstyle{remark}
\newtheorem{rmk}[thm]{Remark}

\numberwithin{equation}{section}

\newcommand{\eps}{\varepsilon}

\newcommand{\DEF}{{:=}}

\newcommand{\Cset}{\mathbb{C}}
\newcommand{\Nset}{\mathbb{N}}

\newcommand{\Rset}{\mathbb{R}}

\newcommand{\Sp}{\mathbb{T}}

\newcommand{\re}{\mathop{\mathrm{Re}}}

\newcommand{\dd}{d} 

\DeclareMathOperator{\sgn}{sgn}

\DeclareMathOperator{\gammafcn}{\Gamma}

\DeclareMathOperator{\digammafcn}{\psi}

\DeclareMathOperator{\sinc}{sinc}

\DeclareMathOperator{\zetafcn}{\zeta}

\newcommand{\Pochhsymb}[2]{{\left(#1\right)_{#2}}}

\allowdisplaybreaks[1]

\title[The Riesz  energy of the $N$-th roots of unity]{The Riesz  energy of the $N$-th roots of unity:  an asymptotic expansion for large $N$} 
\author{ J. S. Brauchart\textasteriskcentered, D. P. Hardin\textdagger, and E. B. Saff\textdaggerdbl } 
\thanks{\noindent \textasteriskcentered  The research of this author was supported, in part,
by the Austrian Science Foundation under grant S9603-N13 and by the U. S. National Science Foundation under grant DMS-0532154  (D. P. Hardin  and E. B. Saff principal investigators).\\
\textdagger The research of this author was supported, in part,
by the U. S. National Science Foundation under grants DMS-0505756 and DMS-0532154. \\
\textdaggerdbl The research of this author  was supported, in part,
by the U. S. National Science Foundation under grants DMS-0532154 and DMS-0603828. 
}

\begin{document}

\address{J. S. Brauchart: Department of Analysis and Computational Number Theory (Math A),
Graz University of Technology, 
Steyrergasse 30, 8010 Graz, Austria
}
\address{D. P. Hardin, E. B. Saff:
Center for Constructive Approximation, 
Department of Mathematics, 
Vanderbilt University, 
Nashville, TN 37240, 
USA }
\email{Johann.Brauchart@Vanderbilt.Edu}
\email{Doug.Hardin@Vanderbilt.Edu}
\email{Edward.B.Saff@Vanderbilt.Edu}

\keywords{Riesz energy, Fekete points, Euler-Maclaurin summation, Logarithmic energy, Riemann zeta function, }
\subjclass[2000]{}

\begin{abstract} 
We derive the complete asymptotic expansion in terms of powers of $N$ for  the Riesz $s$-energy of $N$ equally spaced points on the unit circle  
as $N\to \infty$.   For $s\ge -2$, such points form optimal energy $N$-point configurations  with respect to  the Riesz potential $1/r^{s}$, $s\neq0$, where $r$ is the Euclidean distance between points.  
 By analytic continuation  we deduce the expansion for all complex values of $s$.
  The Riemann zeta function plays an essential role in this asymptotic expansion.  
\end{abstract}

\maketitle

\section{Introduction and statement of results}

The {\em Riesz $s$-energy} ($s\neq0$) of $N$ points $z_1,\dots,z_N$ of the complex plane $\Cset$ is defined by
\begin{equation} \label{Es.def}
  E_s(z_1,\dots,z_N) \DEF \sum_{j\neq k} \left| z_j - z_k \right|^{-s} \DEF \sum_{j=1}^N\sum_{\begin{subarray}{c} k=1 \\ k\neq j \end{subarray}}^N \left| z_j - z_k \right|^{-s}.
\end{equation}
The limiting case as  $s\to 0$ (of the derivative of $E_s$) yields the {\em logarithmic energy}: 
\begin{equation} \label{E0.def}
  E_0(z_1,\dots,z_N) \DEF \sum_{j\neq k} \log\frac{1}{\left| z_j - z_k \right|}.  \end{equation}

For an infinite compact set $A\subset \Cset$ and real $s$, we define the {\em optimal $N$-point $s$-energy of $A$} by
\begin{equation}
\mathcal{E}_s(A,N) \DEF 
\begin{cases} 
\inf\left\{ E_s(z_1,\dots,z_N) \mid z_1,\dots,z_N\in A \right\} & \text{if $s\geq0$,} \\ 
\sup\left\{ E_s(z_1,\dots,z_N) \mid z_1,\dots,z_N\in A \right\} & \text{if $s<0$.} 
\end{cases}
\end{equation}

It can be shown using a convexity (concavity) argument that for $s\ge -1$ and each $N\ge 2$, the $N$-th roots of unity 
$$z_{k,N}:=\exp\{2\pi i (k-1)/N\}, \qquad k=1,\dots,N,$$ form optimal $N$-point $s$-energy configurations for the unit circle $\Sp:=\{z\in \Cset \mid  |z|=1\}$ (cf. \cite{AS74}, \cite{LFTRF}, and \cite{gotz}).  
The optimality of the $N$-th roots of unity for $-2<s<-1$ (as well as all $s\ge -1$) is a special case of a much more general result \cite[Theorem 1.2]{cohnkumar:2007} of Cohn and Kumar.
 That is,   
\begin{equation}
\mathcal{E}_s(\Sp,N) = E_s(z_{1,N},\ldots, z_{N,N}), \qquad   s\ge -2 , \,   N\geq2  .
\end{equation}
   It follows from results of Bj{\"o}rck  \cite{bjoerck:1956} that for $s<-2$ and an even number $N=2K$ of points, optimal energy configurations on $\Sp$ consist  of $K$ points at each end of a diameter of $\Sp$.   The same configuration is optimal for $N$ odd in an asymptotic sense.  Thus, $N$ equally spaced points on $\Sp$ with $N\geq3$ are no longer  optimal   if $s<-2$.  
For $s=-2$, the $N$-th roots of unity are optimal, but so is any configuration of $N$ points on the circle whose centroid is the origin (\cite[Theorem 5]{bjoerck:1956}).

Of particular interest in the study of the discrete energy problem on a compact set $A$  in $\Rset^p$   is the asymptotic expansion of the optimal energy $\mathcal{E}_s(A,N)$ as $N\to\infty$. 
If $A$ has Hausdorff dimension $d$ and $0<s<d$, classical potential theory (cf. Landkof \cite{landkof:72}) gives 
\begin{equation} \label{cal.I.s}
\lim_{N\to\infty} \frac{\mathcal{E}_s(A,N)}{N^2} = \mathcal{I}_s[\mu_{A,s}], 
\end{equation}
where $\mathcal{I}_s[\mu_{A,s}]$ is the energy of the {\em equilibrium measure} $\mu_{A,s}$ on $A$.   More precisely, 
$\mu_{A,s}$ is the unique probability measure supported on $A$ that  minimizes the \emph{$s$-energy}
\begin{equation}
\mathcal{I}_s[\mu] \DEF \int\int \left| z - w \right|^{-s} \dd\mu(z) \dd\mu(w)
\end{equation}
over the class $\mathcal{M}(A)$ of all Borel probability measures $\mu$ supported on $A$. 
That is,
\begin{equation}\label{VsDef}
V_s(A) \DEF \inf\{ \mathcal{I}_s[\mu] \mid \mu\in \mathcal{M}(A) \} = \mathcal{I}_s[\mu_{A,s}].
\end{equation}
An analogous result holds for   $s=0$ with the logarithmic kernel. Equation~\eqref{cal.I.s} is also
valid for the case $-2<s< 0$,
where now $\mu_{A,s}$ maximizes $\mathcal{I}_s[\mu]$ for $\mu\in \mathcal{M}(A)$ (cf.  \cite{bjoerck:1956},  \cite{farkasetal} and \cite{landkof:72}).

Here we focus on the optimal energy problem  for the circle $A=\Sp$ in which case the $s$-equilibrium measure $\mu_{\Sp,s}$ is easily seen to be normalized arclength measure.  For $-2<s<1$, $s\neq 0$,  we have (cf. \cite[2.5.3.1]{prudnikov/brychkov/marichev:86vol1}) 
\begin{equation}\label{V.s.0}
V_s \DEF V_s(\Sp)  = \frac{1}{2\pi}\int_0^{2\pi} |1-e^{i\phi}|^{-s}\dd \phi 
=  2^{-s} \frac{\gammafcn((1-s)/2)}{\sqrt{\pi}\gammafcn(1-s/2)};
\end{equation}
hereafter, we set $V_0=1$ so that $V_s$ is continuous at 0.
Our goal is to  derive a {\em complete} asymptotic expansion of $\mathcal{E}_s(\Sp,N)$ as $N\to \infty$.  For the special case $-2<s<1$, $s\neq 0$, 
the above discussion shows that $V_s N^2$ is the dominant term.  The general results of  \cite{martinez-finkelshtein/maymeskul/rakhmanov/saff} (see the discussion at the end of this section) imply that,  for $A=\Sp$ and   $s=1$, the dominant term 
is $(1/\pi)N^2\log N$, while for $s>1$,  the dominant term in the expansion is $[2\zeta(s)/(2\pi)^s]N^{1+s}$, where $\zeta(s)$ denotes the classical Riemann zeta function.   

We will in fact provide  in Theorems~\ref{main.thm.general} and~\ref{main.thm.exceptional} a   complete  asymptotic expansion  of 
\begin{equation}
\mathcal{L}_s(N):= E_s(z_{1,N},\ldots, z_{N,N}), \qquad N\ge 2, \, s\in \Cset. 
\end{equation}
Three noteworthy features of this expansion are:  (i) the fact that it is valid for complex $s$ in \eqref{Es.def}, (ii) the essential role played by the Riemann zeta function,
and (iii) the unifying feature of these formulas that is provided by use of analytic continuation in the complex variable $s$. 
From an energy standpoint, 
the formulas provide a natural bridge between the potential theoretic case ($-2<s<1$) and 
the hypersingular case ($s\ge 1$).

As we noted above, $\mathcal{L}_s(N)=\mathcal{E}_s(\Sp,N)$ for $s\ge -2$.  
It is easy to see that 
\begin{equation}\label{LsDef}
\mathcal{L}_s(N)=    2^{-s} N \sum_{k=1}^{N-1} \left( \sin \frac{\pi k}{N} \right)^{-s}, \qquad N\ge 2, \, s\in \Cset, \, s\neq 0,
\end{equation}
where $a^{-s}:=e^{-s\ln a}$ for $a>0$ and $s\in \Cset$.

Before we present the results for $s\neq0$, we give a simple formula for the logarithmic energy $\mathcal{L}_0(N)$   which can be obtained in a remarkably easy way by taking the logarithm of both sides of the identity (see, for example, \cite[6.1.1(2)]{prudnikov/brychkov/marichev:86vol1}, \cite[(91.1.3)]{hansen:75}) 
\begin{equation}
\prod_{k=1}^{N-1} \sin\frac{\pi k}{N} = 2^{1-N} N, \qquad N\geq2.
\end{equation}
Indeed, by the symmetry of the points $z_{k,N}$  one gets
$$
\mathcal{L}_0(N) = N \sum_{k=2}^{N} \log\frac{1}{\left| z_{k,N} - z_{1,N} \right|} = N\log\frac{1}{2^{N-1}\prod_{k=1}^{N-1}\sin \frac{\pi k}{N}}=- N \log N.
$$
Alternatively, one can deduce this formula from the relation between the discriminant of the polynomial $p_N(z):=z^N-1$ 
and the resultant of $p_N(z)$ and $p_N'(z)$. 


The Euler-MacLaurin summation formula is an essential tool in obtaining our main results which are stated in terms of the classical
Riemann zeta function 
\begin{equation}
\zetafcn(s) \DEF \sum_{n=1}^\infty \frac{1}{n^s}, \qquad \re s > 1.
\end{equation}
It is well-known that $\zeta(s)$ can be analytically continued to the half-plane $\re s + 2p > 0$ for arbitrary $p$ (cf., for example, \cite[Formula~(1.8)]{ivic:1985_zeta_fcn} \footnote{Note, that the factor $1/(2p+1)!$ is missing in \cite[Formula~(1.8)]{ivic:1985_zeta_fcn}.}) using
\begin{equation}
\begin{split} \label{zeta.half-plane}
\zetafcn(s) &= \frac{1}{s-1} + \frac{1}{2} + \sum_{k=1}^p \frac{B_{2k}}{(2k)!} \Pochhsymb{s}{2k-1} \\ 
&\phantom{=\pm}- \frac{\Pochhsymb{s}{2p+1}}{(2p+1)!} \int_1^\infty C_{2p+1}(x) \, x^{-s-2p-1} \dd x.
\end{split}
\end{equation}
The Pochhammer symbol $(\cdot)_n$, the Bernoulli numbers $B_n$, and the periodic (mod $1$) Bernoulli functions $C_n(x)$ are defined in Section \ref{sec:preliminary}.

 The coefficients $\alpha_n(s)$, $n\ge 0$, defined by the generating function relation
\begin{equation} \label{sinc.power.0}
\left( \frac{\sin \pi z}{\pi z} \right)^{-s} = \sum_{n=0}^\infty \alpha_n(s) z^{2n},
\qquad |z|<1, \ s\in \Cset,
\end{equation}
also play a central role in our results. 
In fact, they may be expressed in terms of {\em generalized Bernoulli polynomials} (also called {\em  N{\o{}}rlund  polynomials}) $B_k^{(\sigma)}(x)$, which are defined by the generating function relation (cf. \cite{roman})
\begin{equation} \label{bernoulli.gen.fcn.rel}
\left( \frac{z}{e^z - 1} \right)^\sigma e^{x z} = \sum_{k=0}^\infty \frac{B_k^{(\sigma)}(x)}{k!} z^k, \qquad   \quad |z|<2\pi,
\end{equation}
by means of
\begin{equation} \label{eq:explicit.a.n.s}
\alpha_n(s) = \frac{(-1)^n B_{2n}^{(s)}(s/2)}{(2n)!} \left( 2 \pi \right)^{2n}, \qquad n=0,1,2,\dots.
\end{equation}

We now present our main results concerning the asymptotic expansion (as $N\to\infty$) of  $\mathcal{L}_s(N)$. 
For this purpose we make use of the fact that  $V_s$ in \eqref{V.s.0} has an analytic  continuation to the
complex plane, more precisely:
\begin{equation} 
\label{V.s}
V_s \DEF  \frac{2^{-s}\gammafcn((1-s)/2)}{\sqrt{\pi}\gammafcn(1-s/2)},  \quad s\in \Cset, s\neq1,3,5, \dots.
\end{equation}

\begin{thm}[general case] \label{main.thm.general}
Let $s\in \Cset$ with $s$ not zero or an odd positive integer and let $p$ be any non-negative integer. Then
\begin{equation} \label{exp1}
\begin{split}
\mathcal{L}_s(N) 
&= V_s \, N^2 + \frac{2}{(2\pi)^s}  \sum_{n=0}^p \alpha_n(s) \zetafcn(s-2n) N^{1+s-2n} \\
&\phantom{=\pm}+ \mathcal{O}_{s,p}(N^{-1+\re s-2p}), \qquad N\to\infty.
\end{split}
\end{equation}
\end{thm}

\begin{rmk}
If $s$ is a non-zero even integer, then $\zetafcn(s-2n)$ can be zero, since the Riemann zeta function $\zetafcn(s)$ has its trivial zeros at $s=-2,-4,-6,\dots$. Consequently, the asymptotic expansion of $\mathcal{L}_s(N)$ terminates after finitely many terms and one gets exact formulas. That is, the expansion \eqref{exp1} simplifies to
\begin{equation} 
\mathcal{L}_{s}(N) = V_{s} N^2, \qquad s=-2,-4,-6,\dots,
\end{equation}
and for $s=2M$, $M=1,2,3,4,\dots$, the expansion \eqref{exp1} becomes 
\begin{equation} \label{exp2xct}
\mathcal{L}_s(N) = \frac{2}{(2\pi)^s}   \sum_{n=0}^M \alpha_n(s) \zetafcn(s-2n) N^{1+s-2n}.
\end{equation}
(The constant $V_s$ vanishes for $s=2,4,6,8,\dots$; see \eqref{V.s}.) 
We further remark that when $s=0$,  the right-hand side of \eqref{exp1} reduces to $N(N-1)$ which is precisely the limit as $s\to 0$ of 
$\mathcal{L}_s(N)$ (see \eqref{LsDef}).
\end{rmk}

\begin{thm}[exceptional case] \label{main.thm.exceptional}
Let $s=2M+1$, $M=0,1,2,3,\dots$, and let $p$ be any integer with $p>M$. Then
\begin{equation}
\begin{split} \label{exp3}
\mathcal{L}_s(N) 
&= \frac{1}{\pi} \frac{\Pochhsymb{1/2}{M}}{2^{2M} M!} N^2 \log N + \left( G_M + \frac{1}{\pi} \frac{\Pochhsymb{1/2}{M}}{2^{2M}M!} \gamma \right) N^2 \\
&\phantom{=\pm}+ \frac{2}{(2\pi)^s}   \sum_{\begin{subarray}{c} n=0, \\ n\neq M \end{subarray}}^p \alpha_n(s) \zetafcn(s-2n) N^{1+s-2n} \\
&\phantom{=\pm}+ \mathcal{O}_{s,p}(N^{-1+s-2p}), \qquad N\to\infty,
\end{split}
\end{equation}
where the constant $G_M$ is given by
\begin{equation*}
G_M = \frac{1}{\pi} \frac{\Pochhsymb{1/2}{M}}{2^{2M} M!}\left[ \frac{\alpha_M^\prime(2M+1)}{\alpha_M(2M+1)} + \frac{1}{2} \digammafcn( M + 1 ) - \frac{1}{2} \digammafcn( M + 1 / 2 ) - \log \pi \right],
\end{equation*}
with the derivative with respect to $s$ of $\alpha_M(s)$ at $s=2M+1$   given by
 \begin{equation*}
\alpha_M^\prime(2M+1) = \sum_{m=0}^{M-1} \alpha_m(2M+1) \frac{\zetafcn(2(M-m))}{M-m},
\end{equation*}
and $\gamma = \lim_{M\to\infty} ( 1 + \frac{1}{2} + \frac{1}{3} + \frac{1}{4} + \cdots + \frac{1}{M} - \log M )$ denotes the Euler-Mascheroni constant. 
\end{thm}

\begin{rmk} \label{rmk:sign} A delicate question is that of the sign of the coefficients in the asymptotic expansion of $\mathcal{L}_s(N)$ as $N\to\infty$. We discuss the coefficient
\begin{equation} \label{c.n.s}
c_n(s) \DEF \frac{2}{(2\pi)^s} \alpha_n(s) \zetafcn(s-2n),
\end{equation}
associated with the power $N^{1+s-2n}$ in \eqref{exp1}. For $s>0$ the sign of $c_n(s)$ is determined by $\zetafcn(s-2n)$, since $\alpha_n(s)>0$ in this case (see Proposition \ref{alphapositive}). 
Hence
\begin{align}
\sgn c_n(s) &= \sgn \zetafcn(s-2n), \qquad s>0. 
\end{align}
(Recall, the Riemann zeta function $\zetafcn(s^\prime)$ is positive for $s^\prime>1$, negative for $-2<s^\prime<1$, and changes its sign when $s^\prime$ moves over one of the trivial zeros of $\zetafcn(s^\prime)$ at $s^\prime=-2,-4,-6,\dots$.) 
\end{rmk}

\begin{rmk}\label{rmk:div}
Of further interest is the asymptotic behavior of the coefficient $c_n(s)$ in \eqref{c.n.s} as $n\to\infty$. 
We already know (cf. Remark \ref{rmk:sign}) that $c_n(2M)=0$ for $M$ an integer and $n>M$. 
For $s>0$ and $s\neq2,4,\dots$, we have  (see Appendix~\ref{Ab})
\begin{equation} \label{c.n.asymptotics}
\limsup_{n\to \infty} |c_n(s)N^{-2n}| = \infty \qquad \text{for    $N\geq2$ fixed.}
\end{equation}
This means that the asymptotic expansion of $\mathcal{L}_s(N)$ as $N\to\infty$ forms a divergent series if $s>0$ and $s\neq 2,4,\dots$.  
\end{rmk}

\bigskip 

We conclude this section by mentioning some related results for optimal Riesz $s$-energy on one-dimensional rectifiable 
curves $A$ in $\Rset^p$.  For such curves,   Mart{\'{\i}}nez-Finkelshtein et al. \cite{martinez-finkelshtein/maymeskul/rakhmanov/saff} obtained the dominant term in the hypersingular case $s\ge 1$.  Namely, they showed
\begin{equation}
\lim_{N\to\infty} \frac{\mathcal{E}_1(A,N)}{N^2\log N} = \frac{2}{|A|}, \qquad \lim_{N\to\infty} \frac{\mathcal{E}_s(A,N)}{N^{1+s}} = \frac{2\zetafcn(s)}{|A|^{s}}, \quad s>1,
\end{equation}
where   $|\cdot|$ denotes here the arc length.  (For results concerning more general sets $A\subset\Rset^p$,   we refer to \cite{hardin/saff:notices,hardin/saff:2005}).  
 Borodachov \cite{borodachov:phd_thesis} derived the next order term for sufficiently smooth closed curves  for $s\geq1$. 
In his formula,   the curvature $\kappa(x)$ of the curve $A$ at $x\in A$ plays a crucial role.  For example, for a simple closed $C^3$ curve (in $\Rset^p$) and $s>3$ there holds \cite[Thm.~V.1.1]{borodachov:phd_thesis}
\begin{equation}
\lim_{N\to\infty} \frac{\mathcal{E}_s(A,N)-(2\zetafcn(s)/|A|^{s})N^{1+s}}{N^{s-1}} = \frac{s \zetafcn(s-2)}{12 |A|^{s-2}} \left\|\kappa \right\|_{L^2(A,\lambda_A)}^2.
\end{equation}
Here $\lambda_A$ denotes the normalized arclength measure supported on $A$. 
Even less is known about the next order term for $0<s<1$ (the dominant term is given by \eqref{cal.I.s}).  
The case of the unit circle analyzed in this paper is the first example where all terms in the asymptotic expansion of the optimal energy are explicitly given.  
 This is possible because one knows precisely the positions of the optimal points and can use their symmetry. 
 
 The outline of the paper is as follows.  In Section 2, we present notation and formulas that are central for the proofs of our main theorems.   In Section 3, we use the Euler-Maclaurin formula to obtain an expansion for $\mathcal{L}_s(N)$ in terms of the
 incomplete zeta function.  In Section 4, we provide proofs of our main theorems.  In Appendix A, we derive an alternate form for the constant $G_M$ appearing in Theorem~\ref{main.thm.exceptional}  while in Appendix B we provide a proof of the assertion  \eqref{c.n.asymptotics} in Remark~\ref{rmk:div}.

\section{Preliminaries}
\label{sec:preliminary}

\subsection{The function $\sinc^{-s} z$ and another expression for $V_s$.}
\label{subsec:sinc}

The normalized `sinc'  function defined by 
\begin{equation*}
\sinc z \DEF
\begin{cases}
(\sin \pi z)/(\pi z) & \text{if $z\neq0,$} \\  
1 & \text{if $z=0$,} 
\end{cases}
\end{equation*}
is an  entire function that is  non-zero  for $|z|<1$ and hence, has a logarithm $g(z)= \log \sinc z$ that is analytic for $|z|<1$ (we choose the branch such that $\log \sinc 0=0$).  The  function  $\sinc^{-s} z \DEF \exp(-s \log \sinc  z)$   is even and analytic on the unit disc $|z|<1$ and  thus has a power series representation of the form
\begin{equation} \label{sinc.power}
\sinc^{-s} z = \sum_{n=0}^\infty \alpha_n(s) z^{2n}, \quad |z|<1,\, s\in \Cset. 
\end{equation}
Since
\begin{equation*}
\sinc^{-s} z = \left( \frac{2\pi i z}{\exp(i \pi z) - \exp(-i \pi z)} \right)^{s} = \left( \frac{2\pi i z}{\exp(i 2 \pi z) - 1} \right)^{s} \exp(i \pi s z),
\end{equation*}
it follows from \eqref{bernoulli.gen.fcn.rel} that the coefficients $\alpha_n(s)$ can be expressed in terms of generalized Bernoulli polynomials (cf. relation \eqref{eq:explicit.a.n.s}).

For $0<y<1$, we  have 
\begin{equation}\label{sin.exp}
\left( \sin \pi y \right)^{-s} = \pi^{-s} y^{-s}  \sinc^{-s} y = \pi^{-s} \sum_{n=0}^\infty \alpha_n(s) y^{2n-s},
\end{equation}
where the convergence is uniform for $y$ in any compact subset of $(0,1)$ (and $s$ fixed). 
Hence we may differentiate term-wise to obtain 
\begin{equation} \label{sin.dif}
\frac{\dd^m}{\dd y^m} \left( \sin \pi x y \right)^{-s} = \frac{(-1)^m}{\pi^s} \sum_{n=0}^\infty \alpha_n(s) \Pochhsymb{s-2n}{m} x^{2n-s} y^{2n-s-m},
\end{equation}
where  $\Pochhsymb{a}{m}$ denotes the {\em Pochhammer symbol} $\Pochhsymb{a}{m}\DEF a(a+1)\cdots(a+m-1)$
and $\Pochhsymb{a}{0}\DEF 1$. 
Similarly, we  construct an antiderivative $A_s(y)$ for $(\sin \pi y)^{-s}$ by integrating term-wise to get
\begin{align} \label{As.1}
A_s(y) &\DEF \frac{1}{\pi^{s}} \sum_{n=0}^\infty \alpha_n(s) \frac{y^{2n+1-s}}{2n+1-s}, \quad s\neq  1, 3, 5, \dots; \\
\intertext{in the exceptional cases $s=2M+1$, $M=0,1,2,\dots$, we have} 
A_s(y) &\DEF \frac{\alpha_M(s)}{\pi^{s}} \log y + \frac{1}{\pi^{s}} \sum_{\begin{subarray}{c} n=0, \\ n\neq M \end{subarray}}^\infty \alpha_n(s) \frac{y^{2n+1-s}}{2n+1-s},\quad s=2M+1. \label{As.2}
\end{align}

By Proposition~\ref{alphapositive} in the Appendix, $\alpha_n(s)$ is a polynomial in $s$ with non-negative coefficients. Therefore, $|\alpha_n(s)| \leq \alpha_n(|s|) \leq \alpha_n(R)$ for $|s|\leq R$ and so, using the convergence of \eqref{sinc.power} at $z=y$, it follows that the  expression for $A_s(y)$ in \eqref{As.1} converges uniformly for $s$ in any compact subset  of $\Cset \setminus \{1, 3, 5, \ldots\}$ for   fixed $0<y<1$.  Hence,  $A_s(y)$ is an analytic function of $s$ for $s\neq 1, 3, 5, \ldots$ for   fixed $0<y<1$.  

When $\re s<1$, the function $(\sin\pi y)^{-s}$ is integrable on $[0,1]$ and,
using the symmetry of $\sin \pi y$ about $y=1/2$, we have (recall \eqref{V.s.0})
\begin{equation*} 
V_s = 2^{1-s}\int_0^{1/2} \left(\sin\pi y\right)^{-s} \dd y = 2^{1-s}A_s(1/2) = \frac{1}{\pi^s} \sum_{n=0}^\infty \frac{ \alpha_n(s)(1/2)^{2n}}{2n-s+1}.
\end{equation*}
Since $V_s$ agrees with $2^{1-s}A_s(1/2)$ for $\re s<1$ and both $V_s$ and  $2^{1-s}A_s(1/2)$ are analytic functions of $s$ for $s\neq 1, 3, 5, \ldots$, they must agree for all such $s$ and so we have
\begin{equation}\label{V.s.2}
V_s = 2^{1-s}A_s(1/2) = \frac{1}{\pi^s} \sum_{n=0}^\infty \frac{ \alpha_n(s)(1/2)^{2n}}{2n-s+1}, \quad s\neq 1,3,5,\ldots.
\end{equation}
In the exceptional cases $s=2M+1$, $M=0,1,2,\ldots$, we define the constant $G_M$ by
\begin{equation} \label{G.M}
\begin{split}
G_M &= 2^{1-s}A_s(1/2) = \frac{\alpha_M(s)}{2^{s-1}\pi^{s}} \log \frac{1}{2} + \frac{1}{\pi^{s}} \sum_{\begin{subarray}{c} n=0, \\ n\neq M \end{subarray}}^\infty \alpha_n(s) \frac{(1/2)^{2n}}{2(n-M)}.
\end{split}
\end{equation}

\subsection{The Euler-MacLaurin formula and incomplete Riemann zeta function}\label{euler-mac:sec}
The numbers $B_k$ in the summation formula
\begin{equation}
\frac{1}{p+1} n^{p+1} + \frac{1}{2} n^p + \sum_{k=2}^p \binom{p}{k-1} \frac{B_k}{k} n^{p-k} = 1^p + 2^p + \cdots + n^p
\end{equation}
together with $B_0\DEF1$, $B_1\DEF-1/2$ are called {\em Bernoulli numbers}. In particular, one has $B_{2k+1}=0$ and $(-1)^{k-1}B_{2k}>0$ for $k=1,2,3,\dots$.  
The {\em Bernoulli polynomials} are defined by
\begin{equation} \label{bernoulli.poly}
B_n(x) \DEF \sum_{k=0}^n \binom{n}{k} B_{n-k} x^k.  \end{equation}
The first two Bernoulli polynomials are then $B_0(x)=1$, $B_1(x)=x-1/2$.
These polynomials satisfy the generating function relation
\begin{equation} \label{gen.fcn.rel}
\frac{t e^{x t}}{e^t - 1} = \sum_{n=0}^\infty  \frac{B_n(x)}{n!}t^n
\end{equation}
and satisfy on the interval $[0,1]$ the following inequalities for $k\geq1$:
\begin{equation} \label{bernoulli.inequ}
\left| B_{2k}(x) \right| \leq \left| B_{2k} \right| \qquad \text{and} \qquad \left| B_{2k+1}(x) \right| \leq \left( 2k+1 \right) \left| B_{2k} \right|. 
\end{equation}
Replacing $x$ by $1-x$ and $t$ by $-t$   leaves the left-hand side of \eqref{gen.fcn.rel} invariant and so we have
\begin{equation}
B_n(1-x) = (-1)^n B_n(x), \qquad  x \in \Rset,\  n\geq 0. \label{symm.rel}
\end{equation}

We recall the Euler-MacLaurin summation formula (cf., for example, \cite{apostol:1999} and \cite{teubner:1996}) which states that for a sufficiently smooth function $f$ (for example, if $f^{(2p+1)}$ is continuous on $[1,n]$)
\begin{equation} \label{EulerSum}
\sum_{k=1}^n f(k) = \int_1^n f(x) \dd x + \frac{1}{2} \left\{ f(1) + f(n) \right\} + S_n,
\end{equation}
where
\begin{equation} \label{EulerSum1}
S_n \DEF \frac{B_2}{2!} f^\prime + \frac{B_4}{4!} f^{(3)} + \cdots + \frac{B_{2p}}{(2p)!} f^{(2p-1)} \Big|_1^n + R_p, \qquad p=2,3,\dots,
\end{equation}
and the remainder term is given by
\begin{equation} \label{EulerSum2}
R_p \DEF \frac{1}{(2p+1)!} \int_1^n f^{(2p+1)}(x) C_{2p+1}(x) \dd x , 
\end{equation}
where  the {\em Bernoulli periodic function} $C_n(x)\DEF B_n(x-[x])$ denotes the Bernoulli polynomial   evaluated at the fractional part of $x$.  Property 
\eqref{symm.rel} implies 
\begin{equation} \label{Csymm.rel}
C_n(N-x) = (-1)^n C_n(x), \qquad x\in \Rset, \ N\in \Nset .
\end{equation}

\begin{defn} \label{def:incomplete}
Let $p$ be a non-negative integer, $y \geq1$, and $s$ a complex number with $s\neq1$. Then we define   
\begin{equation}
\begin{split} \label{zeta.in}
\zetafcn_{y,p}(s) &\DEF \frac{1}{s-1} + \frac{1}{2} + \sum_{k=1}^p \frac{B_{2k}}{(2k)!} \Pochhsymb{s}{2k-1} \\
&\phantom{=\pm}- \frac{\Pochhsymb{s}{2p+1} }{(2p+1)!} \int_1^y C_{2p+1}(x) x^{-s-2p-1} \dd x.
\end{split}
\end{equation}
and we also define the quantity 
\begin{equation}\label{Psi.def}
\Psi_{y,p} \DEF \lim_{s\to 1}\left[\zeta_{y,p}(s)-\frac{1}{s-1}\right]=\frac{1}{2} + \sum_{k=1}^p \frac{B_{2k}}{2k} - \int_1^y C_{2p+1}(x) x^{-2p-2} \dd x. 
\end{equation}
\end{defn}


For $\re s + 2p > 0$,  
using  \eqref{zeta.in},  \eqref{zeta.half-plane}, and \eqref{bernoulli.inequ} we get 
\begin{equation}\label{iz.bnd.1}
\begin{split}
| \zetafcn_{y,p}(s) - \zetafcn(s)|&=\left|\frac{\Pochhsymb{s}{2p+1} }{(2p+1)!} \int_y^\infty C_{2p+1}(x) x^{-s-2p-1} \dd x\right| \\
&\le \frac{|\Pochhsymb{s}{2p+1}B_{2p}|}{(2p)! (\re s +2p)}y^{-\re s -2p}=\mathcal{O}_{s,p}(y^{-\re s -2p}).
\end{split}
\end{equation}
In particular, 
\begin{equation}
\lim_{y\to\infty} \zetafcn_{y,p}(s) = \zetafcn(s), \qquad  \re s+2p>0,
\end{equation}
and so $\zetafcn_{y,p}(s)$ can be considered to be an incomplete Riemann zeta function.  
%
%

\begin{rmk}
The quantity  $\Psi_{N/2,p}$  plays an important role in the expansion of $\mathcal{L}_s(N)$ for the exceptional cases
$s=1, 3, 5, \ldots$.  We note that $\Psi_{y,p}$ approaches the Euler-Mascheroni constant $\gamma$ as
$y\to\infty$, since the Euler-MacLaurin summation formula applied to the function $f(x)=1/x$ for any integer  $ p >1$ gives 
\begin{equation*}
\gamma \DEF \lim_{n\to\infty} \left( \sum_{k=1}^n \frac{1}{k} - \log n \right) =  \frac{1}{2} + \sum_{k=1}^p \frac{B_{2k}}{2k} - \int_1^\infty C_{2p+1}(x) x^{-2 p-2} \dd x = \lim_{y\to\infty} \Psi_{y,p}.
\end{equation*}
\end{rmk}

\section{A series expansion of $\mathcal{L}_s(N)$}
\label{sec:direct}

We first establish a series expansion of $\mathcal{L}_s(N)$.  Theorem~\ref{main.thm.general} will then follow from this expansion.  

\begin{lem} \label{series.general}
 Let $p$ be a non-negative integer and let  $s$ be a complex number. 
 \begin{itemize}
 \item[(a)] If $s\neq0,1,3,5,\dots$, then
\begin{equation} \label{series1}
\mathcal{L}_s(N) = V_s N^2 + \frac{2}{(2\pi)^s}  \sum_{n=0}^\infty \alpha_n(s) \zetafcn_{N/2,p}(s-2n) N^{1+s-2n},
\end{equation}
where $V_s$ is given in \eqref{V.s}
and the numbers $\alpha_n(s)$, $n\geq0$, are defined in \eqref{sinc.power}.

\item[(b)] If $s= 2M+1$, $M=0,1,2,3,\dots$, then
\begin{align} \label{series3}
\mathcal{L}_s(N) 
&= 2^{1-s} \frac{\alpha_M(s)}{\pi^s} N^2 \log N+\left( G_M  + 2^{1-s} \frac{\alpha_M(s)}{\pi^s} \Psi_{N/2,p} \right) N^2      \notag \\
&\phantom{=}+ \frac{2}{(2\pi)^s}   \sum_{\begin{subarray}{c} n=0, \\ n\neq M \end{subarray}}^\infty \alpha_n(s) \zetafcn_{N/2,p}(s-2n) N^{1+s-2n},
\end{align}
where  $G_M$ is defined  in  \eqref{G.M}
and  $\Psi_{N/2,p}$ is defined in \eqref{Psi.def}.
\end{itemize}
\end{lem}

\begin{proof}
For $N=1,2,3,\ldots$, let $f_N(x) \DEF [ \sin (\pi x / N) ]^{-s}$ for  $0 < x < N$.
Then from \eqref{LsDef} we have 
\begin{equation}\label{Lser.0}
\displaystyle \mathcal{L}_s(N)=2^{-s}N\sum_{k=1}^{N-1} f_N(k). 
\end{equation} 
Since $f_N(N-x)=f_N(x)$, we  have  $f_N^{(m)}(N-x)=(-1)^mf_N^{(m)}(x)$. 
   Applying the Euler-MacLaurin summation formula \eqref{EulerSum} and using the 
fact that  $C_{2p+1}(x)  f_N^{(2p+1)}(x)$ is even about $x=N/2$ we obtain
\begin{equation}
\begin{split} \label{red.EulerSum}
\sum_{k=1}^{N-1} f_N(k) &=2 \int_1^{N/2} f_N(x) \dd x + f_N(1) - 2 \sum_{k=1}^p \frac{B_{2k}}{(2k)!} f_N^{(2k-1)}(1) \\
& + \frac{2}{(2p+1)!} \int_1^{N/2} C_{2p+1}(x) \, f_N^{(2p+1)}(x) \dd x. 
\end{split}
\end{equation}

We next find series expansions for each of the terms on the right-hand side of \eqref{red.EulerSum}.  Recalling that $A_s(y)$ is an antiderivative of $(\sin \pi y)^{-s}$ for $0<y<1$ we have 

$$J_N:= 2\int_1^{N/2} f_N(x) \dd x = 2 N \int_{1/N}^{1/2} \frac{\dd y}{\left( \sin \pi y \right)^{s}}  = 2 N \left[ A_s(1/2) - A_s(1/N) \right]. $$
 It then follows from \eqref{As.1} and \eqref{V.s.2} that,  for $s\neq0, 1, 3, 5, \dots$, 
\begin{align}
  J_N=&\ 2^s  N V_s + \frac{2}{\pi^s} \sum_{n=0}^\infty \alpha_n(s) \frac{N^{s-2n}}{s-2n-1},  \label{Lser.1}
\intertext{while for $s=2M+1$, $M=0,1,2,\ldots$, we get from \eqref{As.2} and \eqref{G.M} that}
J_N = &\ 2^s N G_M + 2 \frac{\alpha_M(s)}{\pi^s} N \log N + \frac{2}{\pi^s} \sum_{\begin{subarray}{c} n=0, \\ n\neq M \end{subarray}}^\infty \alpha_n(s) \frac{N^{s-2n}}{s-2n-1}. \label{Lser.1b}
\end{align}
Furthermore, from  \eqref{sin.exp}, we have
\begin{equation}\label{Lser.2}
f_N(1) = \frac{1}{2} \frac{2}{\pi^s} \sum_{n=0}^\infty \alpha_n(s) N^{s-2n}
\end{equation}
and from \eqref{sin.dif}  we get
\begin{equation}\label{Lser.3}
\begin{split}
-2\sum_{k=1}^p \frac{B_{2k}}{(2k)!} f_N^{(2k-1)}(1) = \frac{2}{\pi^s} \sum_{n=0}^\infty \alpha_n(s) N^{s-2n} \sum_{k=1}^p \frac{B_{2k}}{(2k)!} \Pochhsymb{s-2n}{2k-1}.
\end{split}
\end{equation}
Moreover, we have
\begin{align}\label{Lser.4}
\frac{2}{(2p+1)!} \int_1^{N/2}  & C_{2p+1}(x) \, f_N^{(2p+1)}(x) \dd x = - 
\frac{2}{\pi^s}  \sum_{n=0}^\infty \alpha_n(s)N^{s-2n}  \\&\times \frac{\Pochhsymb{s-2n}{2p+1}}{(2p+1)!} \int_1^{N/2} C_{2p+1}(x) x^{2n-s-2p-1}\dd x. \notag
\end{align}

Substituting the expressions   \eqref{Lser.1}, \eqref{Lser.2}, \eqref{Lser.3}, and \eqref{Lser.4} into \eqref{Lser.0} and gathering common terms we have in the case $s\neq1,3,5,\ldots,$
\begin{equation*}
\begin{split}
\mathcal{L}_s (N)
&= N^2 V_s + \frac{2}{(2\pi)^s}   \sum_{n=0}^\infty \alpha_n(s) N^{1+s-2n} \\ 
&\phantom{=\times}\times \Bigg\{ \frac{1}{s-2n-1} + \frac{1}{2} + \sum_{k=1}^p \frac{B_{2k}}{(2k)!} \Pochhsymb{s-2n}{2k-1} \\ 
&\phantom{=\times\pm}- \frac{\Pochhsymb{s-2n}{{(2p+1)}}}{(2p+1)!} \int_1^{N/2} C_{2p+1}(x)x^{-(s-2n)-2p-1} \dd x \Bigg\}, 
\end{split}
\end{equation*}
which, using Definition \ref{def:incomplete}, establishes the general case (a). 

Now suppose $s=2M+1$  for some $M=0,1,2,3,\ldots$.  
Substituting the expressions from \eqref{Lser.1b} to \eqref{Lser.4} into \eqref{Lser.0}, separating out the $n=M$ terms in the expressions from \eqref{Lser.2} to \eqref{Lser.4} and using $\Pochhsymb{s-2M}{k}=k!$, we obtain
\begin{equation*}
\begin{split}
\mathcal{L}_s (N)
&= N^2 G_M + \frac{2\alpha_M(s)}{(2\pi)^s} N^2 \log N + \frac{2}{(2\pi)^s}  \sum_{\begin{subarray}{c} n=0, \\ n\neq M \end{subarray}}^\infty \alpha_n(s) N^{1+s-2n} \\ 
&\phantom{=\times}\times \Bigg\{ \frac{1}{s-2n-1} + \frac{1}{2} + \sum_{k=1}^p \frac{B_{2k}}{(2k)!} \Pochhsymb{s-2n}{2k-1} \\ 
&\phantom{=\times\pm}- \frac{\Pochhsymb{s-2n}{2p+1}}{(2p+1)!} \int_1^{N/2} C_{2p+1}(x) x^{-(s-2n)-2p-1} \dd x \Bigg\} \\
&\phantom{=}+ \frac{2\alpha_M(s)}{(2\pi)^s} N^2 \left[ \frac{1}{2} + \sum_{k=1}^p \frac{B_{2k}}{2k} - \int_1^{N/2} C_{2p+1}(x) x^{-2p-2} \dd x \right], 
\end{split}
\end{equation*}
which, using Definition \ref{def:incomplete}, completes the proof of case (b). 
\end{proof}
 
\section{Proofs of Theorems \ref{main.thm.general} and  \ref{main.thm.exceptional}}
\label{sec:proof.main.thm}

We first prove Theorem \ref{main.thm.general} using Lemma \ref{series.general}. 

\begin{proof}[Proof of Theorem~\ref{main.thm.general}] Let $p$ be an integer $\geq0$ and let $q$ be the smallest positive integer such that $\re s  + 2 q > 2$.
Then by Lemma~\ref{series.general}, we can write
\begin{align}
\mathcal{L}_s(N) 
&=V_s N^2 + \frac{2}{(2\pi)^s}   \sum_{n=0}^p \alpha_n(s) \zetafcn(s-2n) N^{1+s-2n}  \notag\\
&\phantom{=}+ \frac{2}{(2\pi)^s}  \sum_{n=0}^p \alpha_n(s)\left[\zetafcn_{N/2,p+q}(s-2n) -\zetafcn(s-2n)\right]N^{1+s-2n}  \label{tr.1}\\
&\phantom{=}+ \frac{2}{(2\pi)^s}   \sum_{n=p+1}^\infty \alpha_n(s) \zetafcn_{N/2,p+q}(s-2n) N^{1+s-2n}. \label{tr.2}
\end{align}
Using \eqref{iz.bnd.1}, it follows that the expression \eqref{tr.1} is $\mathcal{O}_{s,p}(N^{-1+\re s -2p})$.

To estimate the last term  \eqref{tr.2}, we write it as 
\begin{equation}\label{tr.3}
\frac{2N^{-1+s-2p}}{(2\pi)^s}   \sum_{n=p+1}^\infty \beta_n(s,p,N).
\end{equation}
where 
$$\beta(s,p,N):=\alpha_n(s) \zetafcn_{N/2,p+q}(s-2n) (N/2)^{ -2(n-p-1)}(1/2)^{ 2(n-p-1)}.
$$
Observe that $ (s-2n)_{2p+2q+1}$ is a polynomial in $n$ of degree $2p+2q+1$ and that there is a constant $K_1$ depending only on 
$s$ and $p$ such that\footnote{In the case that $s$ is an even integer, it is possible that the integral in \eqref{incompleteIntegral}
introduces a term of order $\log N$; however, in such a case, $n-p-1$ is necessarily positive and hence its product with $\log N$ is bounded by a constant.}
\begin{equation}\label{incompleteIntegral}
(N/2)^{ -2(n-p-1)}\left|  \int_1^{N/2} C_{2p+2q+1}(x) x^{-s+2n-2p-2q-1} \dd x\right|\le K_1
\end{equation}
for all $N\ge 2$ and $n\ge p+1$.  
Then, it follows that $$ \zetafcn_{N/2,p+q}(s-2n) (N/2)^{ -2(n-p-1)}=\mathcal{O}_{s,p}(n^{2p+2q+1})$$ for $n\ge p+1$ and $N\ge 2$. 
Finally, since $\limsup_{n\to \infty}  |\alpha_n(s)|^{1/n}\le 1$ (cf. \eqref{sinc.power}), 
we deduce that there is some constant $K_2$ depending only on $s$ and $p$ such that
$$
 |\beta(s,p,N)| \le K_2 (1/3)^n \quad 
$$  
for all $N\ge 2$ and $n\ge p+1$.   Consequently, \eqref{tr.2} is of order $\mathcal{O}(N^{-1+\re s -2p})$ which combined with the estimate for \eqref{tr.1} completes the proof of Theorem \ref{main.thm.general}.
\end{proof}

Similarly we can prove Theorem~\ref{main.thm.exceptional}  dealing with the exceptional cases $s=1,3,5,\dots$ .
Note that an explicit expression of the constant $G_M$ is given in \eqref{const.G.M} of the Appendix.

\appendix

\section{The constant $G_M$}
The following alternative form  for $V_s$  
\begin{equation}\label{V.s.alt}
V_s
= \frac{2^{-s}\gammafcn(s/2)\tan ({\pi s}/{2})}{\sqrt{\pi}\gammafcn((1+s)/2)},  \quad s\neq1,3,5, \dots,
\end{equation}
  can be obtained from \eqref{V.s} using the identity $\Gamma(z)\Gamma(1-z)=\pi/\sin \pi z$.  Equating the two expressions  \eqref{V.s.alt} and \eqref{V.s.2} and separating out the $n=M$ term ($M=0,1,2,3,\ldots$), we obtain for 
  $s \neq 1,3,5,\ldots$ the identity 
  \begin{equation}
  \label{G.M.aux1}
\frac{1}{\pi^s} \sum_{\begin{subarray}{c} n=0, \\ n\neq M \end{subarray}}^\infty \frac{ \alpha_n(s)(1/2)^{2n}}{2n-s+1} 
 =  \frac{2^{-s}\gammafcn(s/2)\tan ({\pi s}/{2})}{\sqrt{\pi}\gammafcn((1+s)/2)}  - \frac{\alpha_M(s)}{\pi^s} \frac{(1/2)^{2M}}{2M-s+1} .
\end{equation}
The left-hand side in \eqref{G.M.aux1}, and therefore also its right-hand side, is analytic at $s=2M+1$ (cf. the discussion in Section \ref{subsec:sinc} of the antiderivative $A_s(y)$). Set $s(\eps)=2M+1+2\eps$, $|\eps|<1$, where $M$ is a fixed non-negative integer, and define the auxiliary function
\begin{equation}
H_M(\eps) \DEF \frac{1}{\pi^{s(\eps)}} \sum_{\begin{subarray}{c} n=0, \\ n\neq M \end{subarray}}^\infty \frac{ \alpha_n(s(\eps))(1/2)^{2n}}{2(n-M-\eps)}.
\end{equation}
By continuity of $H_M(\eps)$ at $\eps=0$ and \eqref{G.M}, we obtain
\begin{equation} \label{G.M.aux4}
\lim_{\eps\to0} H_M(\eps) = G_M + \frac{\alpha_M(2M+1)}{2^{2M} \pi^{2M+1}} \log 2.
\end{equation}
Since
$
\tan({\pi s}/{2}) = \tan \pi \left( M + 1/2 + \eps \right) = \tan \pi \left( 1/2 + \eps \right) = - \cot \pi \eps,
$
it follows
\begin{equation} \label{G.M.aux3}
H_M(\eps)  = \frac{2^{-2M}}{2\eps\pi^{s(\eps)}}   \Bigg\{ \alpha_M(s(\eps)) - \pi^{2M} \frac{\gammafcn(M+1/2+\eps)}{\sqrt{\pi}\gammafcn(M+1+\eps)} \left( \frac{\pi}{2} \right)^{2\eps} \left( \pi \eps \cot \pi \eps \right) \Bigg\}.
\end{equation}
From the existence of the limit \eqref{G.M.aux4} and relation \eqref{G.M.aux3} we immediately get
\begin{equation} \label{G.M.aux5}
\alpha_M(2M+1) = \pi^{2M} \frac{\Pochhsymb{1/2}{M}}{M!}.
\end{equation}
On the other hand, from \eqref{G.M.aux3}, \eqref{G.M.aux5}, and the fact that $\alpha_n(s(\eps))$ are polynomials in $\eps$
(see Proposition~\ref{alphapositive}), we obtain
\begin{align*}
&\lim_{\eps\to0} H_M(\eps) 
= \frac{2^{-2M}}{\pi^{2M+1}} \lim_{\eps\to0} \frac{\alpha_M(2M+1+2\eps)-\alpha_M(2M+1)}{2\eps} \\
&\phantom{=}+ \frac{2^{-2M}}{2\pi} \frac{\Pochhsymb{1/2}{M}}{M!} \lim_{\eps\to0} \frac{1}{\eps} \Bigg[ 1 - \frac{\gammafcn(M+1/2+\eps)\gammafcn(M+1)}{\gammafcn(M+1/2)\gammafcn(M+1+\eps)} \left( \frac{\pi}{2} \right)^{2\eps} \left( \pi \eps \cot \pi \eps \right) \Bigg]
\end{align*}
The square bracketed expression has the following expansion in $\eps$,
\begin{equation*}
\left( 2 \log \frac{2}{\pi} - \digammafcn( M + 1/2 ) + \digammafcn( M + 1 ) \right) \eps + \mathcal{O}(\eps^2), \quad \text{as $\eps\to0$.}
\end{equation*}
All together this gives
\begin{equation}
\begin{split} \label{G.M.aux7}
&\lim_{\eps\to0} H_M(\eps) = \frac{2^{-2M}}{\pi} \frac{\alpha_M^\prime(2M+1)}{\pi^{2M}} \\
&\phantom{=\pm}- \frac{2^{-2M}}{\pi} \frac{\Pochhsymb{1/2}{M}}{M!} \left[ \frac{1}{2} \digammafcn( M + 1/2 ) - \frac{1}{2} \digammafcn( M + 1 ) + \log \frac{\pi}{2} \right].
\end{split}
\end{equation}
Combining \eqref{G.M.aux4} and \eqref{G.M.aux7} and using \eqref{G.M.aux5}, we have
\begin{equation}
\begin{split}
G_M &= \frac{2^{-2M}}{\pi} \frac{\alpha_M^\prime(2M+1)}{\pi^{2M}} \\
&\phantom{=}- \frac{2^{-2M}}{\pi} \frac{\Pochhsymb{1/2}{M}}{M!} \left[ \frac{1}{2} \digammafcn( M + 1/2 ) - \frac{1}{2} \digammafcn( M + 1 ) + \log \pi \right]
\end{split}
\end{equation}
or equivalently
\begin{equation} \label{const.G.M}
G_M = \frac{2^{-2M}}{\pi} \frac{\Pochhsymb{1/2}{M}}{M!} \left[ \frac{\alpha_M^\prime(2M+1)}{\alpha_M(2M+1)} + \frac{1}{2} \digammafcn( M + 1 ) - \frac{1}{2} \digammafcn( M + 1 / 2 ) - \log \pi \right].
\end{equation}
The first derivative of $\alpha_M(s)$ with respect to $s$ can be expressed in terms of coefficients $\alpha_n(s)$ with $n<M$.

\begin{lem}
The coefficients $\alpha_n(s)$  in \eqref{sinc.power.0} satisfy
\begin{equation} \label{alpha.n.prime}
\alpha_0^\prime(s) = 0, \qquad \alpha_n^\prime(s) = \sum_{m=0}^{n-1} \alpha_m(s) \frac{\zetafcn(2(n-m))}{n-m}, \quad n\geq1,
\end{equation}
for all $s\in\Cset$.
\end{lem}

\begin{proof}
Differentiating both sides of   \eqref{sinc.power.0} we have
\begin{equation} \label{G.M.aux10}
\left( - \log \sinc z \right)\sinc^{-s} z  = \sum_{n=0}^\infty \alpha_n^\prime(s) z^{2n}, \qquad |z|<1, \ s \in \Cset.
\end{equation}
Taking the logarithm of both sides of the infinite product
\begin{equation}
\frac{\sin \pi z}{\pi z} = \prod_{k=1}^\infty \left( 1 - \frac{z^2}{k^2} \right), \qquad z\in\Cset,
\end{equation}
and substituting the Maclaurin series of $\log (1-z)$,  we obtain the series expansion
\begin{equation} \label{log.sinc.expansion}
- \log \sinc z = \sum_{m=1}^\infty \frac{\zetafcn(2m)}{m} z^{2m}, \qquad |z|<1.
\end{equation}
Substituting the expansions of $\sinc^{-s}z$ and \eqref{log.sinc.expansion} into \eqref{G.M.aux10} and gathering common terms, a comparison of the coefficients of $z^{2n}$ on both sides in \eqref{G.M.aux10} gives \eqref{alpha.n.prime}.
\end{proof}

\begin{prop}\label{alphapositive}
For each $n=0,1,2,\ldots$, the coefficient $\alpha_n(s) $ is a polynomial of degree $n$ in $s$ with non-negative coefficients. 
\end{prop}
\begin{proof}
The assertion follows by induction on $n$.  Simply integrate \eqref{alpha.n.prime} and use the fact that
$\alpha_0(s)=1$ and that $\alpha_m(0)=0$ for $m\ge 1$.  
\end{proof}
\section{Proof of \eqref{c.n.asymptotics}}\label{Ab}

Let $s\neq0,\pm2,\pm4,\dots$. Using the well-known functional equation for the Riemann zeta function, we have 
\begin{equation}
c_n(s) = (-1)^n \frac{2}{\pi} \left(\sin\frac{\pi s}{2}\right) \frac{\alpha_n(s)}{(2\pi)^{2n}} \gammafcn(2n+1-s) \zetafcn(2n+1-s).
\end{equation}
Assume to the contrary that $|c_n(s)N^{-2n}|$ remains bounded, that is, since $\zetafcn(2n+1-s)\to1$ as $n\to\infty$, there exists a constant $c>0$ such that
\begin{equation*} 
|\alpha_n(s)|\leq c \frac{(2\pi N)^{2n}}{\gammafcn(2n+1-s)} \leq c \left( \frac{2\pi e N}{2n+1-s} \right)^{2n} \left(\frac{e}{2n+1-s}\right)^{-s}
\end{equation*}
for all $n\geq n_0$. (We used $(x/e)^{x-1}<\gammafcn(x)<(x/2)^{x-1}$, $x\geq2$.) Hence,
\begin{equation*}
\limsup\limits_{n\to\infty} \sqrt[2n]{|\alpha_{n}(s)|} = 0.
\end{equation*}
It follows that the series $\sum_{n=0}^\infty \alpha_n(s) z^{2n}$ is an entire function. 
But this series also represents the function $[(\sin \pi z)/(\pi z)]^{-s}$, which has a singularity at $z=1$. 
\hfill $\Box$

\def\cprime{$'$}
\providecommand{\bysame}{\leavevmode\hbox to3em{\hrulefill}\thinspace}
\providecommand{\MR}{\relax\ifhmode\unskip\space\fi MR }
\providecommand{\MRhref}[2]{%
  \href{http://www.ams.org/mathscinet-getitem?mr=#1}{#2}
}
\providecommand{\href}[2]{#2}

%

\end{document}